\documentclass[aps,showpacs,twocolumn]{revtex4}
\usepackage[dvips]{graphicx}
\usepackage{amssymb}
\newcommand{\ped}[1]{\ensuremath{_{\rm #1}}}

\begin{document}

\title{Pressure dependence of critical temperature in $MgB_{2}$ and two bands
Eliashberg theory}

\author{G.A. Ummarino}\email{E-mail:giovanni.ummarino@infm.polito.it}

\affiliation{ Dipartimento di Fisica, Politecnico di Torino, Corso
Duca degli Abruzzi 24, 10129 Torino, Italy} \affiliation{INFM-
LAMIA, Corso Perrone 24, 16152 Genova, Italy}

\begin{abstract}
The variation of the superconducting critical temperature $T_{c}$
as a function of the pressure $p$ in the magnesium diboride
$MgB_{2}$ has been studied in the framework of two-bands
Eliashberg theory and traditional phonon coupling mechanism. I
have solved the two-bands Eliashberg equations using
first-principle calculations or simple assumptions for the
variation, with the pressure, of the relevant physical quantities.
I have found that the experimental $T_{c}$ versus $p$ curve can be
fitted very well and information can be obtained on the dependence
of the electron-phonon interaction matrix $<I^{2}>$ by pressure.
The pressure dependence of the superconductive gaps
$\Delta_{\sigma}$ and $\Delta_{\pi}$ is also predicted.
\end{abstract}

\maketitle

In the last few years, there is an noticeable increase of the
study of superconductivity in many elements under pressure
\cite{rev}, such as sulphur ($T_{c}=17$ K), oxygen ($T_{c}=0.5$
K), carbon in nanotube ($T_{c}=15$ K) and diamond forms
($T_{c}=4$) K, a non-magnetic state of iron ($T_{c}=1$ K), and the
light elements lithium ($T_{c}=20$ K) and boron ($T_{c}=11$ K).
The application of external pressure to superconductors can drive
the compounds towards or away from lattice instabilities by
varying the principal parameters determining the superconducting
properties (the electronic density of states at the Fermi energy,
the characteristic phonon frequency, and the electron-phonon
coupling constant), and it can be used to tune the $T_{c}$ and the
superconducting properties. Almost all of the superconducting
metallic materials, unlike the previous simple elements, show a
decrease of $T_{c}$ with pressure. This negative pressure
coefficient was attributed to the volume dependence of the
electronic density of states at the Fermi energy and of the
effective interaction between the electrons mediated by the
electron-phonon coupling. Measurements of the influence of
pressure on the transition temperature and critical field yield
information on the interaction causing the superconductivity.
Indeed, the pressure would seem to be a variables whose effects
might be capable of immediate theoretical interpretation. The
binary alloy $MgB_{2}$, superconductor \cite{Nagamatsu}, at
ambient pressure, at $T=40$ K has, under pressure, a behaviour
similar to metallic materials. The magnesium diboride has
stimulated intense investigation, both from the theoretical and
the experimental point of view. Now the electronic structure of
$MgB_{2}$ is well understood and the Fermi surface consists of two
three-dimensional sheets, from the $\pi$ bonding and antibonding
bands, and two nearly cylindrical sheets from the two-dimensional
$\sigma$ bands \cite{bandstructure}. There is a large difference
in the electron-phonon coupling on different Fermi surface sheets
and this fact leads to a multiband description of
superconductivity. Theory indicates that the strongest coupling is
realized for the near-zone center in-plane optical phonon
($E_{2g}$ symmetry) related to vibration of the B atoms
\cite{rev}. The superconductivity in $MgB_{2}$ has been deeply
studied in the past three years and so also the effect of pressure
on the superconductive properties. The effect of pressure on the
superconducting properties of $MgB_{2}$ has been studied by
several groups. All groups observed a decrease of $T_{c}$ with
increasing pressure \cite{press1,pressexp} and I want show that
this decrease can be very well explained in the framework of the
two bands Eliashberg theory. In the following I will refer to the
paper of A.F.Goncharov \cite{pressexp} because in there are
present both measurement of the variation of critical temperature
and of phonon mode by means of Raman measurement, with the
pressure and so I mainly refer to these experimental data. In fact
only in this work there are all input parameters necessary to my
model.

Let us start from the generalization of the Eliashberg theory
\cite{Eliashberg,Marsiglio} for systems with two bands
\cite{Kresin}, that has already been used with success to study
the MgB$_{2}$ and related systems
\cite{Brinkman,Golubov,Mazin0,Choi,John1,John2}. To obtain the
gaps and the critical temperature within the $s$-wave, two-band
Eliashberg model one has to solve four coupled integral equations
for the gaps $\Delta_{i}(i\omega_{n})$ and the renormalization
functions $Z_{i}(i\omega_{n})$:
 \begin{eqnarray}
\omega_{n}Z_{i}(i\omega_{n})&=&\omega_{n}+\pi
T\sum_{m,j}\Lambda_{ij}(i\omega_{n}-i\omega_{m})N^{j}_{Z}(i\omega_{m})+\nonumber\\
& & +\sum_{j}\Gamma^{ij}N^{j}_{Z}(i\omega_{n})\label{eq:EE1}
\end{eqnarray}
\begin{eqnarray}
Z_{i}(i\omega_{n})\Delta_{i}(i\omega_{n})&=&\pi
T\sum_{m,j}[\Lambda_{ij}(i\omega_{n}-i\omega_{m})-\mu^{*}_{ij}(\omega_{c})]\cdot\nonumber\\
& &
\hspace{-1.5cm}\cdot\theta(|\omega_{c}|-\omega_{m})N^{j}_{\Delta}(i\omega_{m})+\sum_{j}%
\Gamma^{ij}N^{j}_{\Delta}(i\omega_{n}) \label{eq:EE2}
\end{eqnarray}
where $i,j$ are band indices, $\theta$ is the Heaviside function,
$\omega_{c}$ is a cutoff energy, $\Gamma^{ij}$ is the non-magnetic
impurity scattering rate in the Born approximation and:
\begin{equation}
\Lambda_{ij}(i\omega_{n}-i\omega_{m})=\int_{0}^{+\infty}\frac{d\omega
\alpha^{2}_{ij}F(\omega)}{(\omega_{n}-\omega_{m})^{2}+\omega^{2}}
\end{equation}
\begin{equation}
N^{j}_{\Delta}(i\omega_{m})=\frac{\Delta_{j}(i\omega_{m})Z_{j}(i\omega_{m})}%
{\sqrt{\omega^{2}_{m}Z_{j}^{2}(i\omega_{m})+\Delta^{2}_{j}(i\omega_{m})Z_{j}^{2}(i\omega_{m})}}
\end{equation}
\begin{equation}
N^{j}_{Z}(i\omega_{m})=\frac{\omega_{m}Z_{j}(i\omega_{m})}%
{\sqrt{\omega^{2}_{m}Z_{j}^{2}(i\omega_{m})+\Delta^{2}_{j}(i\omega_{m})Z_{j}^{2}(i\omega_{m})}}
\end{equation}
where $\omega\ped{n}=\pi T(2n-1)$ and $n, m=0,\pm 1,\pm 2...$.

The solution of Eqs. 1,2 requires as input: i) the four (but only
three independent\cite{Kresin}) electron-phonon spectral functions
$\alpha^{2}_{ij}(\omega)F(\omega)$; ii) the four (but only three
independent\cite{Kresin}) elements of the Coulomb pseudopotential
matrix $\mu^{*}(\omega\ped{c})$.
%

Let's start with the four spectral functions
$\alpha^{2}_{ij}(\omega)F(\omega)$, that were calculated in ref. 11
(see Fig. 1).

For simplicity, I will assume that the shape of the
$\alpha^{2}_{ij}F(\omega,p)$ functions does not change with the
pressure, and I will only rescale them with the electron-phonon
coupling constants $\lambda_{ij}$:
\begin{equation}
\alpha^{2}_{ij}F(\omega,p)= \frac{\lambda_{ij}(p)}{\lambda_{ij}(0)}
\alpha^{2}_{ij}F(\omega,0)\label{eq:a2F}
\end{equation}
Let me remind the definition of electron-phonon coupling constant
\cite{lambda,lambdaI2,Grimvall}:
\begin{equation}
\lambda=\sum_{q,i}\frac{\gamma_{i}(q)}{\pi N\ped{N}(E\ped{F})
\Omega_{i}^{2}(q)}
 \label{eq:lambda}
\end{equation}
\begin{figure}[!]
 \begin{center}
 \includegraphics[keepaspectratio, width=0.8\columnwidth]{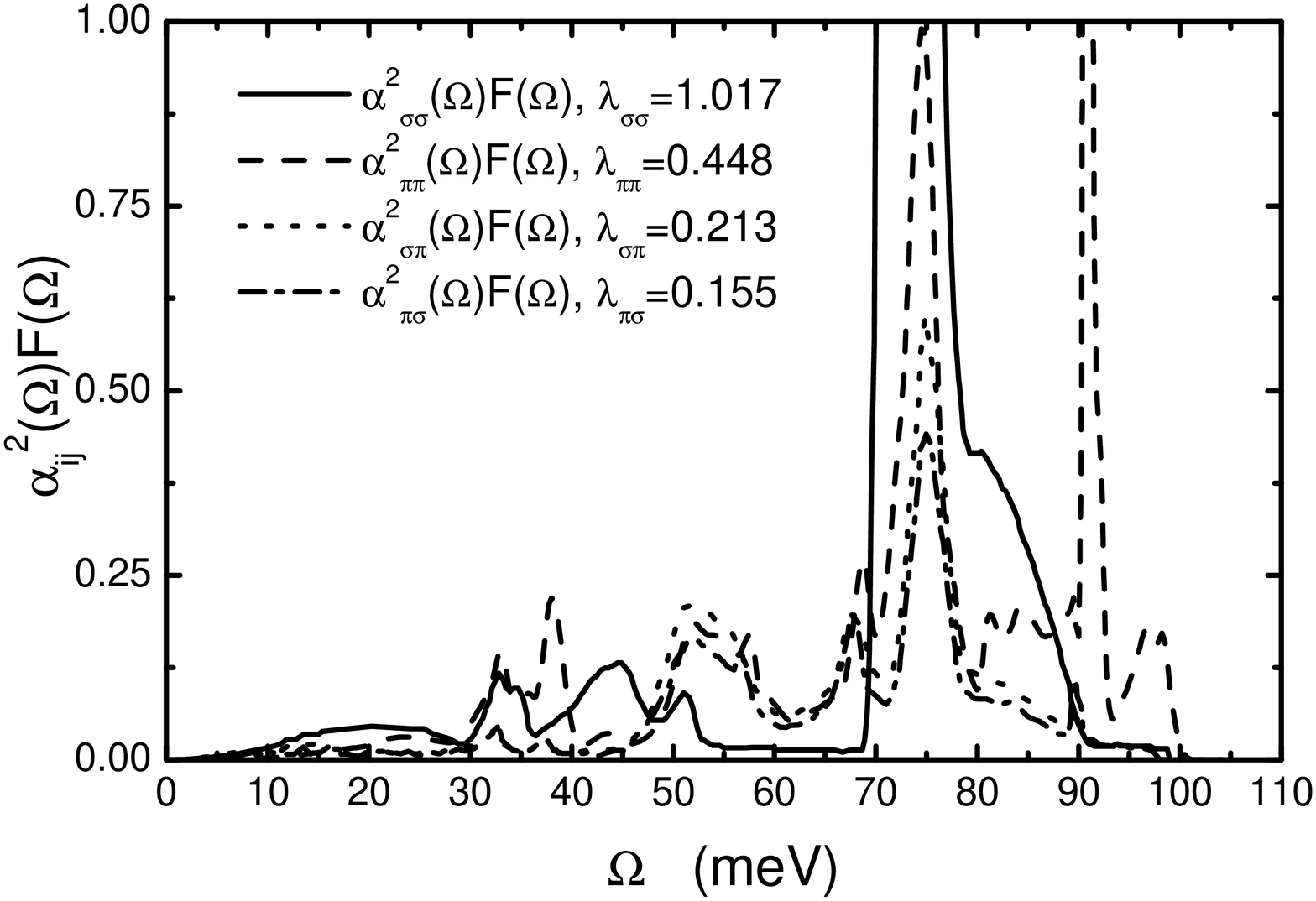}
 \end{center}
 \caption{The spectral functions of the two-band model for the $MgB_{2}$:
$\sigma\sigma$ (solid line), $\pi\pi$ (dashed line), $\sigma\pi$
(dotted line) and $\pi\sigma$ (dashed dotted line), taken from
ref. 11.}
 \end{figure}
\begin{figure}[t]
 \begin{center}
 \includegraphics[keepaspectratio, width=0.8\columnwidth]{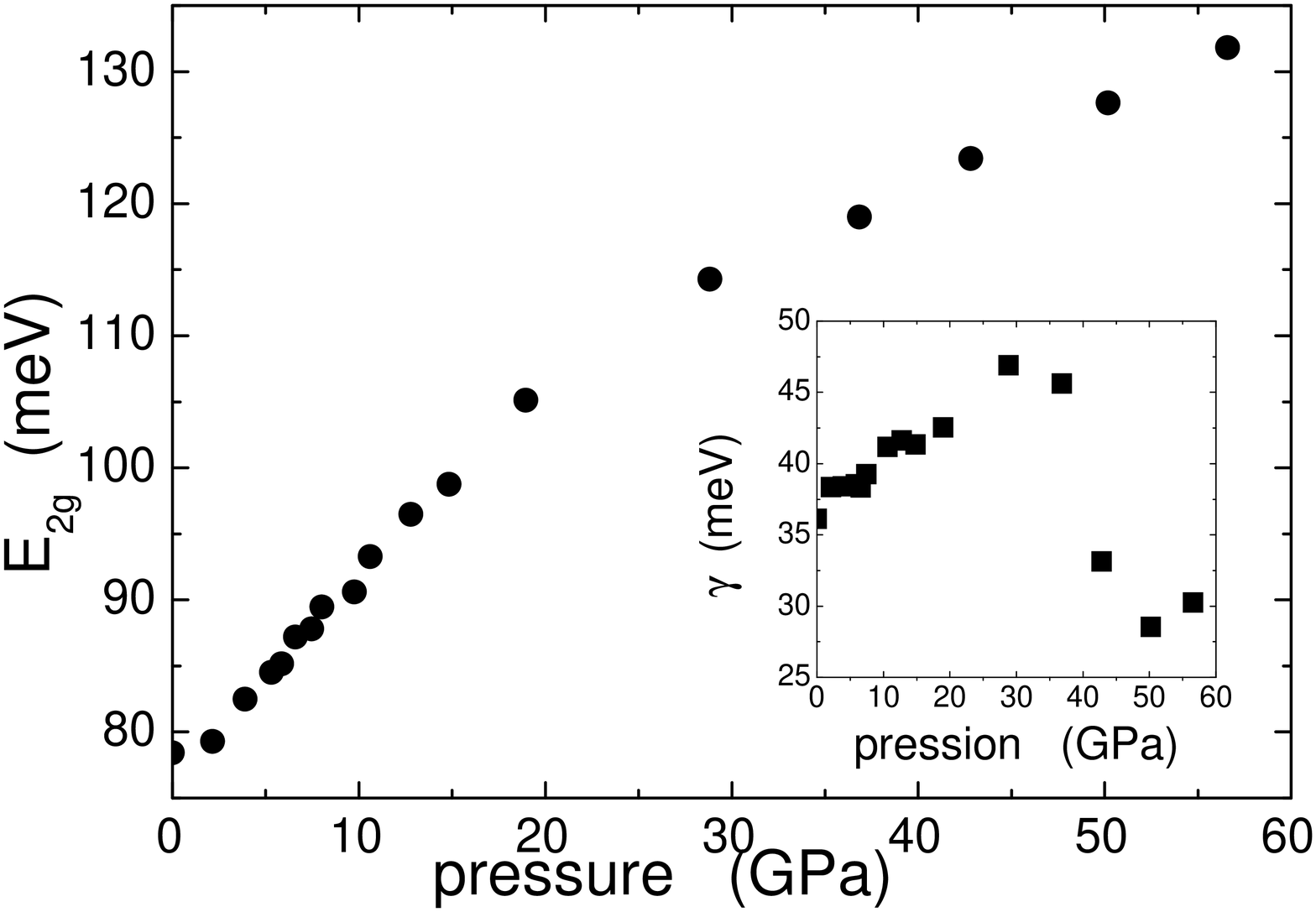}
 \end{center}
 \caption{Experimental energy of phonon $E_{2g}$ of $MgB_{2}$ as a function of the pressure, see ref. 5.
In the insert the experimental Raman linewidth as a function of
the pressure, see always ref. 5.}\label{fig:2}
 \end{figure}
\begin{figure}[!]
 \begin{center}
 \includegraphics[keepaspectratio, width=0.8\columnwidth]{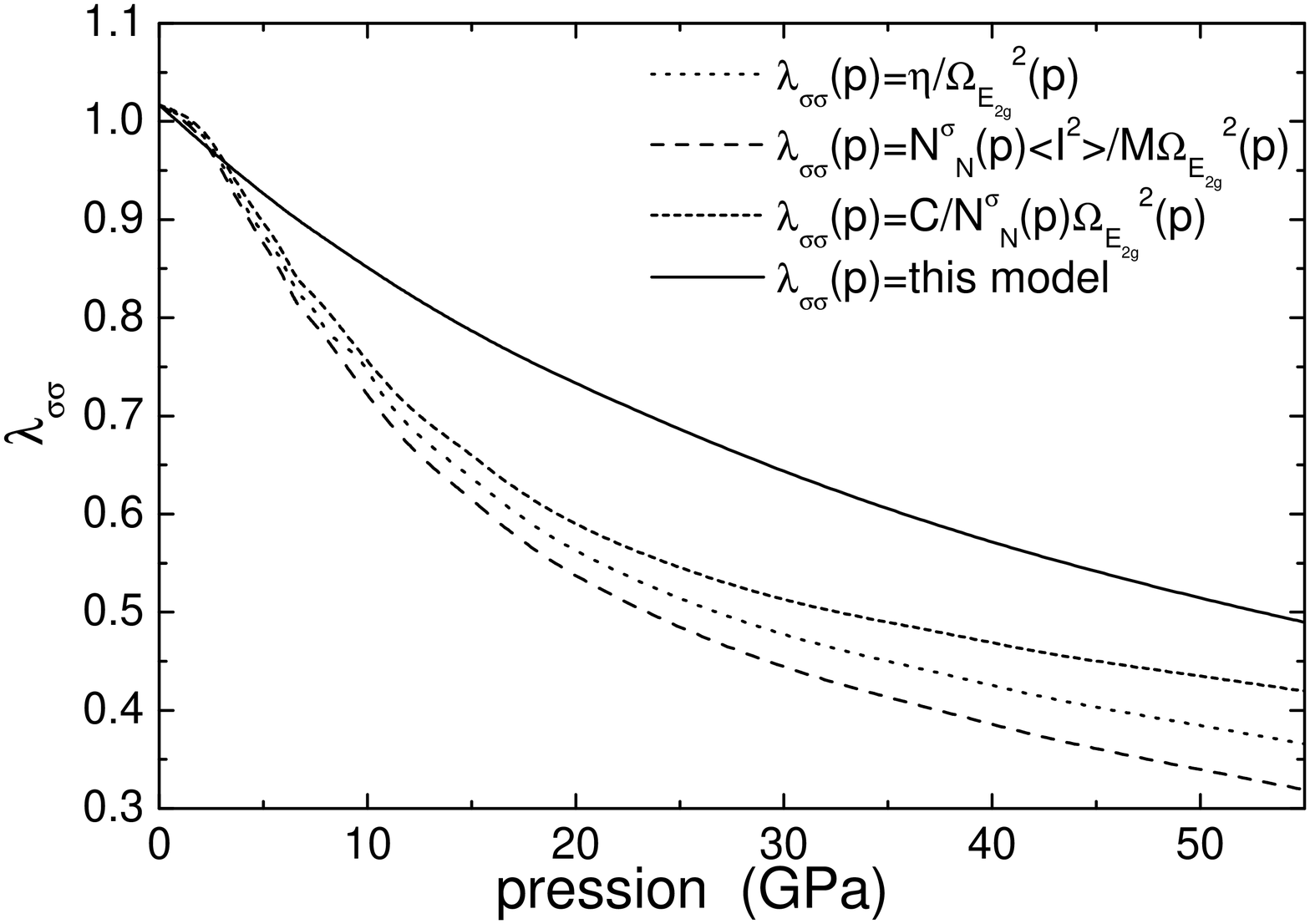}
 \end{center}
 \caption{Calculated electron-phonon coupling constant as a function of
pressure in the four different cases:
$\lambda_{\sigma\sigma}(p)=\eta/\Omega^{2}_{E_{2g}}(p)$ (dotted
line),
$\lambda_{\sigma\sigma}(p)=N\ped{N}^{\sigma}(E\ped{F},p)<I^{2}>/2M
\Omega^{2}_{E_{2g}}(p)$ (dashed line),
$\lambda_{\sigma\sigma}(p)=C/(N\ped{N}^{\sigma}(E\ped{F},p)
\Omega^{2}_{E_{2g}}(p)$) (short dashed line) and
$\lambda_{\sigma\sigma}(p)=[\gamma(p)/\Omega^{2}_{E_{2g}}(p)+\pi\lambda_{\sigma\sigma}(0)N\ped{N}^{\sigma}(E\ped{F},0)-\gamma(0)/\Omega^{2}_{E_{2g}}(0)]/\pi
N\ped{N}^{\sigma}(E\ped{F},p)$ (solid line).}
 \end{figure}
where $\gamma_{i}(q)$ is the phonon linewidth which is the width in
energy of a phonon of momentum $q$, mode index $i$ and energy
$\Omega_{i}(q)$ and $N\ped{N}(E\ped{F})$ is the normal density of
states at the Fermi level. The frequency $\Omega_{i}$ can be
identified with the frequency of the B-B bond-stretching phonon mode
(the $E\ped{2g}$ mode), that has been recently measured as a
function of pressure \cite{pressexp}, and is reported in Fig. 2. In
the insert of Fig. 2 the experimental Raman linewidth that here is
used as phonon linewidth, is shown. Since this mode couples strongly
with the holes on top of the $\sigma$ band, from eq.~\ref{eq:lambda}
I will have for $\lambda\ped{\sigma \sigma}$ (which gives the most
important contribution to superconductivity in our system):
\begin{equation}
\lambda\ped{\sigma\sigma}(p) =\frac{1}{\pi N^{\sigma}
\ped{N}(E\ped{F},p)}
[\frac{\gamma\ped{E\ped{2g}}(p)}{\Omega\ped{E\ped{2g}}(p)}+\sum_{q,i}\frac{\gamma\ped{i}(q)}{\Omega\ped{i}(q)}]
\label{eq:lambda_ss0}
\end{equation}
\begin{equation}
\lambda\ped{\sigma\sigma}(p) =\frac{1}{\pi N^{\sigma}
\ped{N}(E\ped{F},p)}
[\frac{\gamma\ped{E\ped{2g}}(p)}{\Omega\ped{E\ped{2g}}(p)}+C(0)]
\label{eq:lambda_ss1}
\end{equation}
where
\begin{equation}
C(0)=\pi\lambda\ped{\sigma\sigma}(0)N^{\sigma} \ped{N}(E\ped{F},0)-
\frac{\gamma\ped{E\ped{2g}}(0)}{\Omega\ped{E\ped{2g}}(0)}
\label{eq:lambda_ss2}
\end{equation}
When the lattice parameters of $MgB_{2}$ are modified by chemical
substitutions the normal density of states, at the Fermi level, in
the $\pi$-band changes relatively little \cite{Mass} and so I
assume that, in the first approximation,
\begin{equation}
N\ped{N}^{\pi}(E\ped{F},p)=N\ped{N}^{\pi}(E\ped{F},0)
 \label{eq:lambdaNp}
\end{equation}
and
\begin{equation}
N^{\sigma} \ped{N}(E\ped{F},p)=N^{\sigma} \ped{N}(E\ped{F},0)+p
\frac{\partial N^{\sigma} \ped{N}(E\ped{F},p)}{\partial p}|_{p=0}
 \label{eq:lambdaNs}
\end{equation}
I use the values calculated in ref. 10: $N^{\sigma}
\ped{N}(E\ped{F},0)=0.30061$ $(eV unit cell)^{-1}$ and $N^{\pi}
\ped{N}(E\ped{F},0)=0.40359$ $(eV unit cell)^{-1}$ for the
$MgB_{2}$. So $\frac{\partial N^{\sigma}
\ped{N}(E\ped{F},p)}{\partial p}|_{p=0}$ is the only true free
parameter of the model. In this way, I assume that the change in
the frequency of the $E\ped{2g}$ phonon affects the coupling
constant, while I neglect its influence on the shape of the
electron-phonon spectral function \cite{John1}. For the other
coupling constants, I will instead assume for simplicity
\begin{equation}
\hspace{-1mm}\forall (i,j) \neq (\sigma,\sigma) \hspace{5mm}
\lambda\ped{ij}(p)
=\frac{N\ped{N}^{j}(E\ped{F},p)}{N\ped{N}^{j}(E\ped{F},0)}
\lambda\ped{ij}(0)\label{eq:lambda_ij}
\end{equation}
with \cite{Brinkman,Golubov} $\lambda_{\sigma\sigma}(0)=1.017$,
$\lambda_{\pi\pi}(0)$=0.448, $\lambda_{\sigma\pi}(0)$=0.213 and
$\lambda_{\pi\sigma}(0)$=0.155. At the end, in this approximate
model of electron-phonon coupling constants only
$\lambda_{\sigma\sigma}$ and $\lambda_{\pi\sigma}$ change with the
pressure. This fact is in agreement with the results of ref. 20
where the authors find that $\lambda_{\pi\pi}$ is almost constant.
Fig. 3 shows the calculated electron-phonon coupling constant
$\lambda_{\sigma\sigma}$ as a function of the pressure.
%

%
As far as the Coulomb pseudopotential is concerned, let us start
from its expression in pure $MgB_2$
\cite{Brinkman,Golubov,Dolgov}:
\begin{eqnarray}
\hspace{2mm}\mu^{*}(p)= \left| \begin{array}{cc}%
\mu^{*}\ped{\sigma \sigma} & \mu^{*}\ped{\sigma \pi}\\
\mu^{*}\ped{\pi \sigma} & \mu^{*}\ped{\pi \pi}
\end{array} \right| =  \nonumber \\
= \mu(\omega_{c})N^{tot} \ped{N}(E\ped{F},p)
\left| \begin{array}{cc}%
\frac{2.23}{N\ped{N}^{\sigma}(E\ped{F},p)} &
\frac{1}{N\ped{N}^{\sigma}(E\ped{F},p)}\\ & \\
\frac{1}{N\ped{N}^{\pi}(E\ped{F},p)} &
\frac{2.48}{N\ped{N}^{\pi}(E\ped{F},p)}
\end{array} \right| \label{eq:mu}
\end{eqnarray}
where $\mu(\omega\ped{c})$ is a free parameter and $N^{tot}
\ped{N}(E\ped{F},p)$ is the total normal density of states at the
Fermi level. The numbers 2.23 and 2.48 in the Coulomb matrix have
been calculated for the $MgB_{2}$ in ambient pressure but, as a
first approximation, I will suppose them not to depend on the
pressure. In this way, the elements of the Coulomb pseudopotential
matrix, $\mu^{*}\ped{ij}$, depend on the pressure only through the
densities of states at the Fermi level. Now I can fix the cut-off
energy (e.g., $\omega_{c}=700$ meV) so as to reduce the number of
adjustable parameters to two: the prefactor in the Coulomb
pseudopotential, $\mu(\omega_{c})$ and $\frac{\partial N^{\sigma}
\ped{N}(E\ped{F},p)}{\partial p}|_{p=0}$. For having $T_{c}=40.2$
K I fix $\mu(\omega_{c})$ equal to 0.00315 so there is only more a
free parameter for fitting the experimental critical temperature
as a function of pressure.
 \begin{figure}[!]
 \begin{center}
 \includegraphics[keepaspectratio, width=\columnwidth]{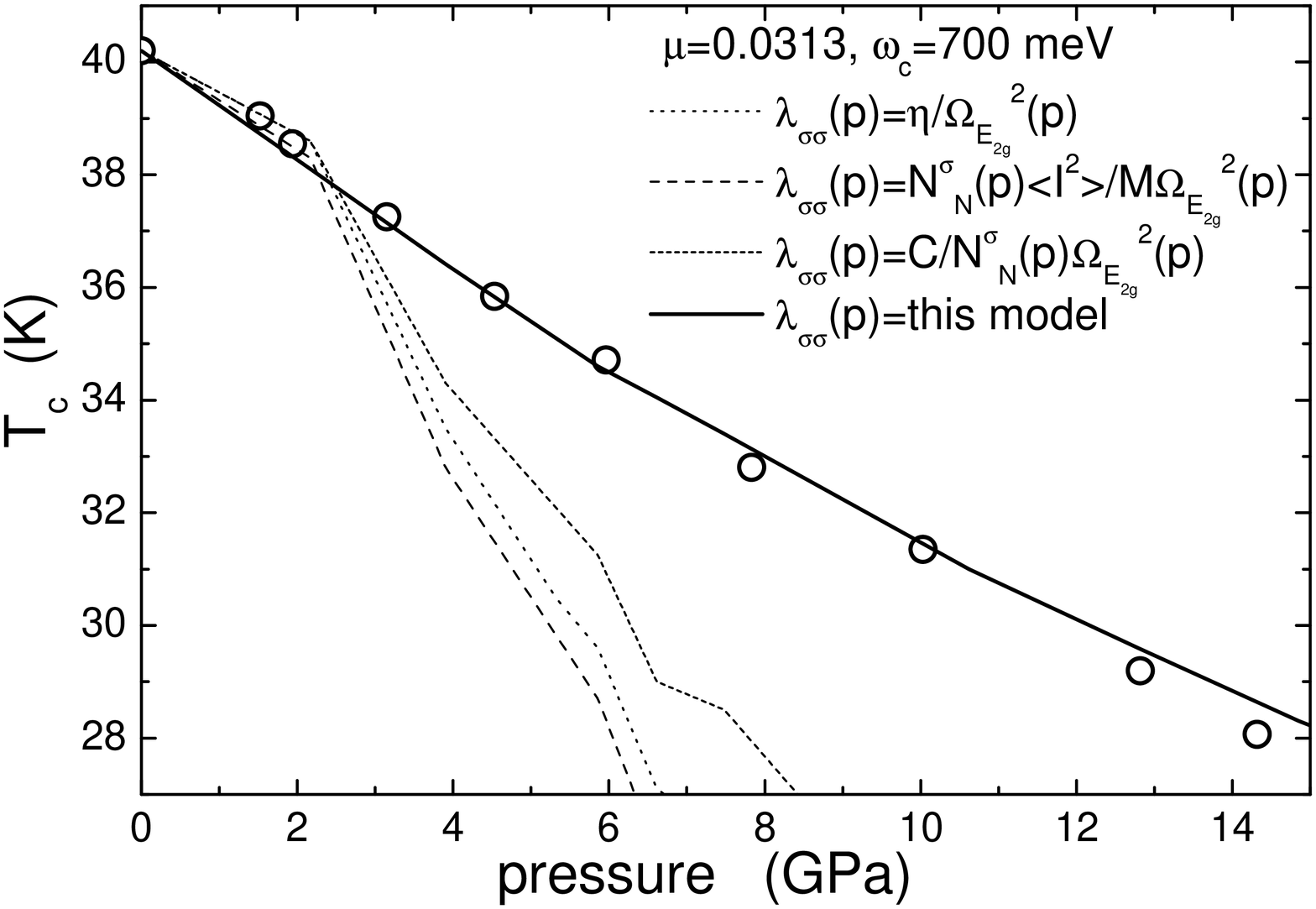}
 \end{center}
  \caption{The measured critical temperature $T_{c}$ as function of  the pressure (open circles)
  and theoretical fits obtained by different assumptions on the
  electron-phonon coupling constant $\lambda_{\sigma\sigma}(p)=\eta/\Omega^{2}_{E_{2g}}(p)$ (dotted line),
$\lambda_{\sigma\sigma}(p)=N\ped{N}^{\sigma}(E\ped{F},p)<I^{2}>/2M
\Omega^{2}_{E_{2g}}(p)$ (dashed line),
$\lambda_{\sigma\sigma}(p)=C/(N\ped{N}^{\sigma}(E\ped{F},p)
\Omega^{2}_{E_{2g}}(p)$) (short dashed line) and
$\lambda_{\sigma\sigma}(p)=[\gamma(p)/\Omega^{2}_{E_{2g}}(p)+\pi\lambda_{\sigma\sigma}(0)
N\ped{N}^{\sigma}(E\ped{F},0)-\gamma(0)/\Omega^{2}_{E_{2g}}(0)]/\pi
N\ped{N}^{\sigma}(E\ped{F},p)$ (solid line).}
 \end{figure}
Before of examining the fit of experimental data with the my
$\lambda_{\sigma\sigma}$ model I can check the other possible
choices for the electron-phonon coupling constants. All cases are
shown in Fig. 3.
The first and simpler possibility is similar to the case of several
transition metals \cite{hop}
\begin{equation}
\lambda(p)=\frac{\eta}{\Omega^{2}_{E_{2g}}(p)}
\end{equation}

where $\eta$ is a constant \cite{Grimvall,I1} and so

\begin{equation}
\lambda_{\sigma\sigma}(p)=\frac{\Omega^{2}_{E_{2g}}(0)}{\Omega^{2}_{E_{2g}}(p)}\lambda_{\sigma\sigma}(0)
\end{equation}

The result is in very poor agreement with experimental data (see
dotted line in Fig. 4).

The second possibility is of that the effect of the pressure is
similar to chemical substitutions as $Al$ and $C$ and so I assume
that \cite{John1}
\begin{equation}
\lambda(p)=\frac{N\ped{N}(E\ped{F},p)<I^{2}>}{M
\Omega^{2}_{E_{2g}}(p)}
\end{equation}
where $M$ is the ion mass \cite{Grimvall} and $<I^{2}>$ does not
depend from the pressure.
 Consequently
\begin{equation}
\lambda_{\sigma\sigma}(p)=\frac{N^{\sigma}
\ped{N}(E\ped{F},p)\Omega^{2}_{E_{2g}}(0)}{ N^{\sigma}
\ped{N}(E\ped{F},0)\Omega^{2}_{E_{2g}}(p)}
\end{equation}
 $\frac{\partial
N^{\sigma} \ped{N}(E\ped{F},p)}{\partial p}|_{p=0}=-0.003$ $(eV
GPa)^{-1}$. Also in this case the result is in very poor agreement
with experimental data (see dashed line in Fig. 4).

The last possibility is suggested by recent band-structure
calculations that show $MgB_{2}$ is a traditional $sp$ metal
superconductor \cite{bandstructure}. The pressure dependence of
$I$ has long been an interesting issue in the research of pressure
effects in simple $sp$ metals \cite{Ko28}. Ziman's calculation of
the electron-phonon interaction led to $<I^{2}>\propto
N\ped{N}(E\ped{F})^{-2}$, at least in the limit of long
wavelengths \cite{Ko30}. So I find
\begin{equation}
\lambda_{\sigma\sigma}(p)=\frac{N^{\sigma}
\ped{N}(E\ped{F},0)\Omega^{2}_{E_{2g}}(0)}{N^{\sigma}
\ped{N}(E\ped{F},p)\Omega^{2}_{E_{2g}}(p)}
\end{equation}
and $\frac{\partial N^{\sigma} \ped{N}(E\ped{F},p)}{\partial
p}|_{p=0}=-0.0007$ $(eV GPa)^{-1}$. As in the previous cases the
result is in very poor agreement with experimental data (see short
dashed line in Fig. 4). Now I can see that my simple model is the
alone that explains the experimental critical temperatures because
other possible models for electron-phonon coupling constant
$\lambda_{\sigma\sigma}$ are incompatible with experimental data.
 \begin{figure}[t]
 \begin{center}
 \includegraphics[keepaspectratio, width=0.8\columnwidth]{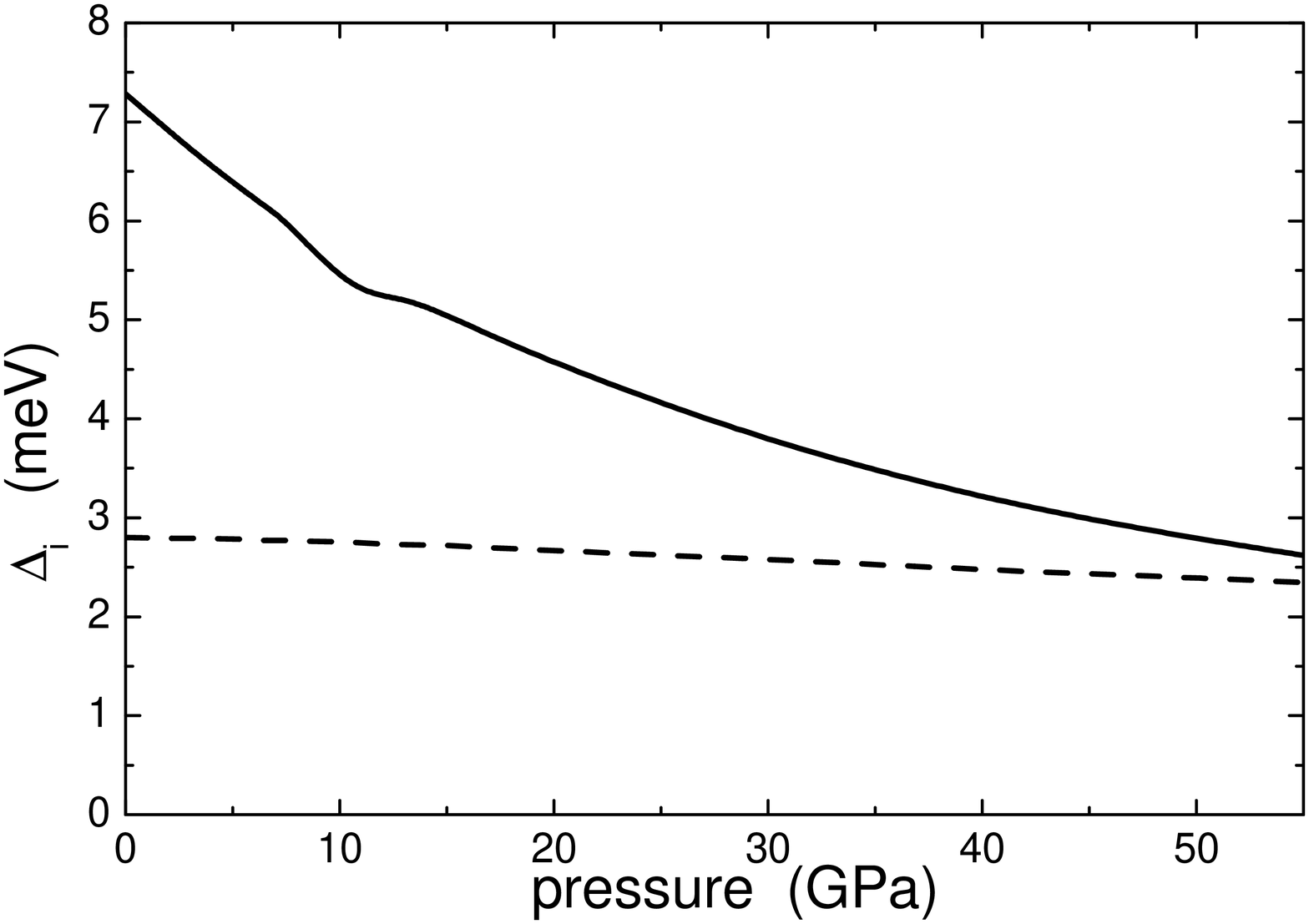}
 \end{center}
 \caption{The calculated values of the gap, at $T=4.2$ K, $\Delta_{\sigma}$ (solid line) and $\Delta_{\pi}$ (dashed line) as a function of the pressure.}
 \end{figure}
 \begin{figure}[t]
 \begin{center}
 \includegraphics[keepaspectratio, width=0.8\columnwidth]{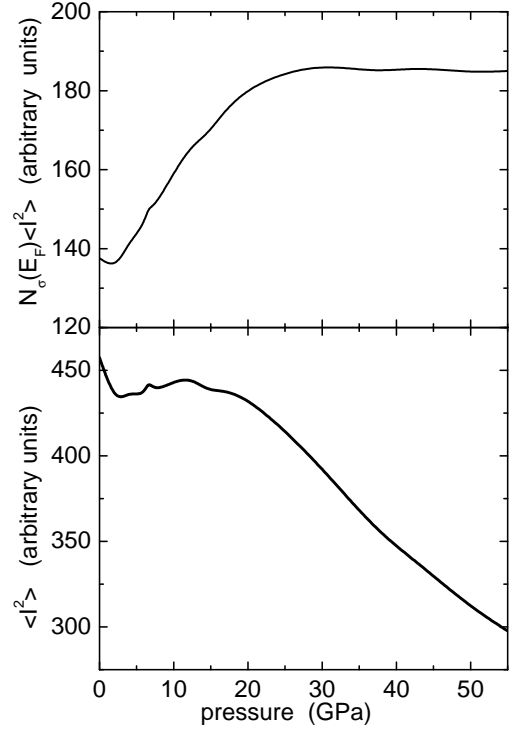}
 \end{center}
  \caption{Upper panel: the calculated value of $N\ped{N}^{\sigma}(E\ped{F})<I^{2}>$ of the $\sigma$-band
  as a function of pressure; lower panel: the calculated value of $<I^{2}>$ of the $\sigma$-band
  as a function of pressure.}
 \end{figure}
\begin{figure}[t]
 \begin{center}
 \includegraphics[keepaspectratio, width=0.8\columnwidth]{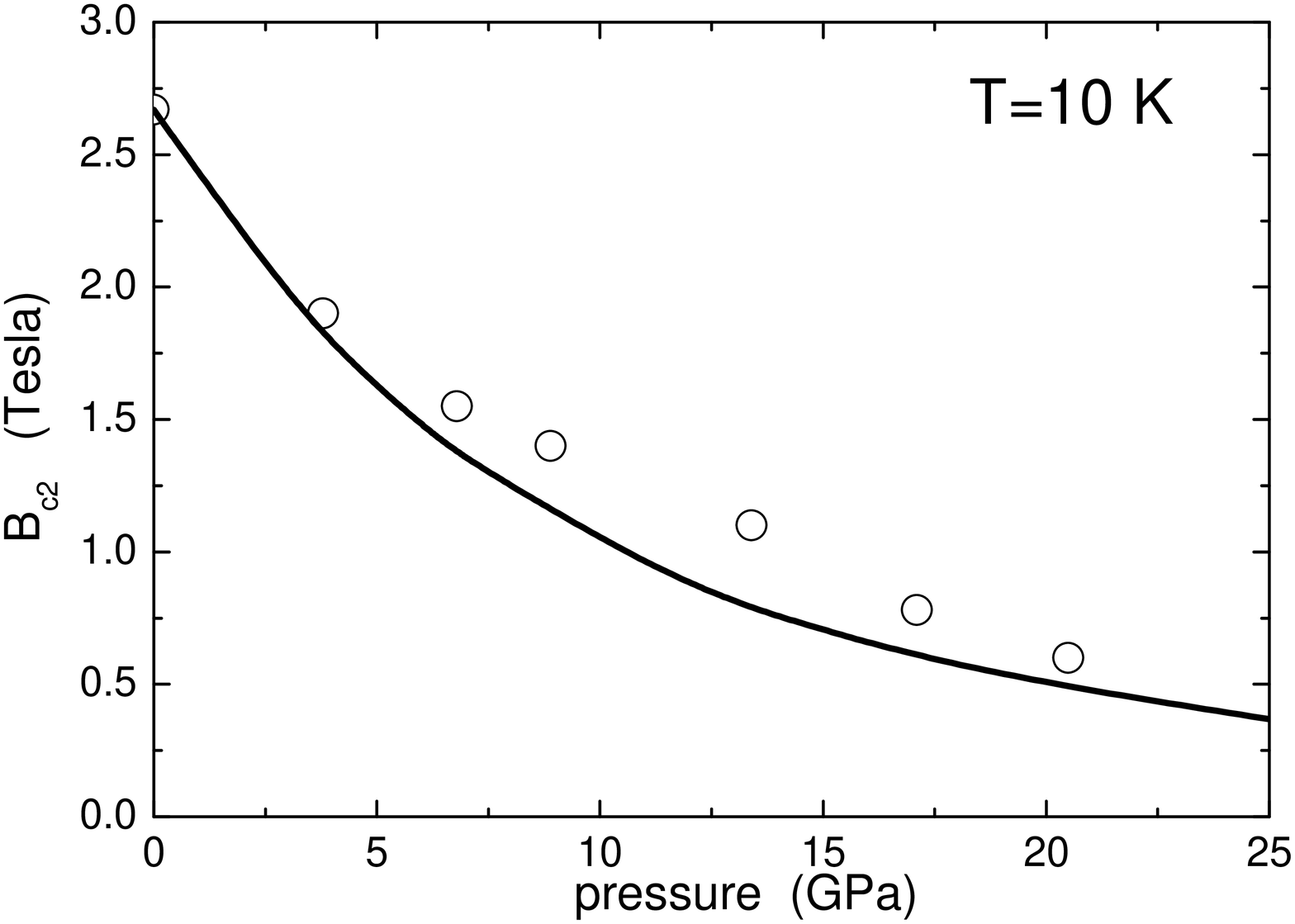}
 \end{center}
  \caption{The measured upper critical field (open circles), at T=10 K, from ref. 20, and the theoretical (solid line).}
 \end{figure}
I obtain the best fit of experimental data (solid line in Fig. 4)
with $\frac{\partial N^{\sigma} \ped{N}(E\ped{F},p)}{\partial
p}|_{p=0}=0.00584$ $(eV GPa)^{-1}$. The fact that $N^{\sigma}
\ped{N}(E\ped{F})$ increases with the pressure is in agreement
with theoretical calculations \cite{Xi}.
The only free parameter of this model is $\frac{\partial
N^{\sigma} \ped{N}(E\ped{F},p)}{\partial p}|_{p=0}$ and so when I
have fixed the optimal value from the $T_{c}$ fit I can calculate,
in principle, all other physical quantities. In Fig. 5 the
theoretical dependence of the $\sigma$ and $\pi$ gaps from the
pressure is shown.

Now from the following equality
\begin{equation}
\frac{N^{\sigma}\ped{N}(E\ped{F},p)<I^{2}(p)>}{M
\Omega\ped{E\ped{2g}}^{2}(p)}=\frac{1}{\pi N^{\sigma}
\ped{N}(E\ped{F},p)}
[\frac{\gamma\ped{E\ped{2g}}(p)}{\Omega\ped{E\ped{2g}}(p)}+\sum_{q,i}\frac{\gamma\ped{i}(q)}{\Omega\ped{i}(q)}]
\end{equation}
it is possible determine the dependence of
$N^{\sigma}\ped{N}(E\ped{F})<I^{2}>$ and of $<I^{2}>$ from the
pressure as it is shown in Fig 6. It can see that, after $\simeq
25$ GPa the Hopfield parameter \cite{Ko30}
$N^{\sigma}\ped{N}(E\ped{F})<I^{2}>$ is almost constant.

At the end it is possible to use this model for explaining the
experimental upper critical field \cite{Sud} in function of
pressure without free parameters. For the sake of completeness, I
give here the linearized gap equations under magnetic field, for a
superconductor in the clean limit (negligible impurity
scattering), as can be found in ref. 20. In the following,
$v_{Fj}$ is the Fermi velocity of band j, and $H_{c2}$ is the
upper critical field:
 \begin{eqnarray}
\omega_{n}Z_{i}(i\omega_{n})&=&\omega_{n}+\pi
T\sum_{m,j}\Lambda_{ij}(i\omega_{n}-i\omega_{m})sign(\omega_{m})\nonumber\\
\end{eqnarray}
\begin{eqnarray}
Z_{i}(i\omega_{n})\Delta_{i}(i\omega_{n})&=&\pi
T\sum_{m,j}[\Lambda_{ij}(i\omega_{n}-i\omega_{m})-\mu^{*}_{ij}(\omega_{c})]\cdot\nonumber\\
& &
\hspace{-1.5cm}\cdot\theta(|\omega_{c}|-\omega_{m})\chi_{j}(i\omega_{m})Z_{j}(i\omega_{m})\Delta_{j}(i\omega_{m})\nonumber\\
\end{eqnarray}
\begin{eqnarray}
\chi_{j}(i\omega_{m})&=&(2/\sqrt{\beta_{j}})\int^{+\infty}_{0}dq\exp(-q^{2})\cdot\nonumber\\
& &
\hspace{-1.5cm}\cdot tan^{-1}[\frac{q\sqrt{\beta_{j}}}{|\omega_{m}Z_{j}(i\omega_{m})|+i\mu_{B}H_{c2}sign(\omega_{m})}]\nonumber\\
\label{eq:EE2}
\end{eqnarray}
with $\beta_{j}=\pi H_{c2} v_{Fj}^{2}/(2\Phi_{0})$.
In these equations the bare Fermi velocities are the input
parameters $v_{Fj}=v^{*}_{Fj}\cdot \sum_{i}(1+\lambda_{ji})$ and
are functions of $p$. For don't having free parameters I assume
that, as in the free electron gas, $v_{Fj}\propto
N^{j}\ped{N}(E\ped{F})$ and so
$v_{Fj}(p)=v_{Fj}(0)N^{j}\ped{N}(E\ped{F},p)/N^{j}\ped{N}(E\ped{F},0)$.
For obtaining exactly the upper critical field of $MgB_{2}$, in
ambient pressure, I find $v^{*}_{F\sigma}(0)=3.6\cdot10^{5}$ m/s
and $v^{*}_{F\pi}(0)=5.35\cdot10^{5}$ m/s in very good agreement
with the calculus of Brinkman et al \cite{Brinkman}. In Fig. 7 is
shown the fit (solid line) of experimental values, from ref. 26,
of $H_{c2}$ at $T=10$ K. The fit isn't so good because, may be,
the approximation of the free electron gas is too strong.

 Finally I conclude by summarizing the main points of this paper. I
have fitted the experimental critical temperatures as a function
of pressure in the framework of two bands Eliashberg theory with
only a free parameter. The result is very good. After I have
calculated other physical quantities can will be compared with
future measurement (for example superconductive gaps from
tunneling curves) and I can affirm that the $MgB_{2}$ under
pressure is, as the same materials in ambient pressure, a moderate
coupling two-band phononic systems well described by two-bands
Eliashberg theory.

Many thanks are due to A. Calzolari for useful discussions.

 \end{document}